\documentclass{pasj00}
\usepackage{epsf}
\SetRunningHead{C.Hikage, A.Taruya, and Y. Suto}
{Biasing and Genus Statistics of Dark Matter Halos}
\newcommand{\deltah}{{\delta_{\rm halo}}}
\newcommand{\deltam}{{\delta_{\rm mass}}}

\newcommand{\bardeltah}{{\overline{\delta_{\rm halo}}}}

\newcommand{\sigmamm}{{\sigma_{\rm mm}}}
\newcommand{\sigmahh}{{\sigma_{\rm hh}}}
\newcommand{\sigmahm}{{\sigma_{\rm hm}}}

\newcommand{\nus}{{\nu_{\rm \sigma}}}
\newcommand{\nuf}{{\nu_{\rm f}}}

\newcommand{\gmaxLN}{{g_{\rm\scriptscriptstyle MAX, LN}}}

\newcommand{\epsilonh}{{s_{\rm halo}}}

\newcommand{\RG}{{\rm\scriptscriptstyle RG}}
\newcommand{\LN}{{\rm\scriptscriptstyle LN}}
\newcommand{\PT}{{\rm\scriptscriptstyle PT}}
\Received{2002/2/9}
\Accepted{2002/12/18}
\begin{document}
\title{Biasing and Genus Statistics of Dark Matter Halos \\
in the Hubble Volume Simulation}
\author{%
Chiaki \textsc{Hikage}, Atsushi \textsc{Taruya} and Yasushi \textsc{Suto}}
\affil{Department of Physics, School of Science, The University of Tokyo, 
Tokyo 113-0033}
\email{hikage@utap.phys.s.u-tokyo.ac.jp, ataruya@utap.phys.s.u-tokyo.ac.jp,
suto@phys.s.u-tokyo.ac.jp}

\maketitle

\begin{abstract}
We present a numerical analysis of genus statistics 
for dark matter halo catalogs from the Hubble volume simulation. 
The huge box-size of the Hubble volume simulation enables us
to carry out a reliable statistical analysis of the biasing
properties of halos at a Gaussian smoothing scale of $R_{\rm G}\ge 30h^{-1}$Mpc
with a cluster-mass scale of between $7\times 10^{13}h^{-1}M_\odot$ and
$6\times 10^{15}h^{-1}M_\odot$.
A detailed comparison of the genus for dark matter halos with that for the mass
distribution shows that the non-Gaussianity induced by the halo biasing
is comparable to that by nonlinear gravitational evolution, and both 
the shape and the amplitude of the genus are almost insensitive to the halo mass
at $R_{\rm G}\ge 30h^{-1}$Mpc.
In order to characterize the biasing effect on the genus, we 
apply a perturbative formula developed by \citet{M1994}.
We find that the perturbative formula well describes the simulated halo genus 
at $R_{\rm G}\ge 50h^{-1}$Mpc. 
The result indicates that the biasing effect on the halo genus is well
approximated by nonlinear deterministic biasing
up to the second-order term in the mass density fluctuation.
The two parameters describing the linear and quadratic terms 
in the nonlinear biasing accurately specify the genus for galaxy clusters.
\end{abstract}

\section{Introduction \label{sec:intro}}

The genus statistics characterizes the topological nature of 
a density field. In contrast to the more conventional 
two-point statistics, such as the two-point correlation functions 
or the power spectrum, the genus is sensitive to the phase information 
of a density field; therefore, the genus provides complementary 
information of the present cosmic structure. 
The genus has an analytic formula in a density field described by random-Gaussian
statistics. Since Gott et al. (1986) proposed the genus of the large-scale
structure as a test of the hypothesis that the initial perturbations 
are Gaussian, the genus has been evaluated for various galaxy or galaxy cluster 
catalogs (e.g., Gott et al. 1989; Park et al. 1992; 
Moore et al. 1992; Rhoads et al. 1994; Vogeley et al. 1994; 
Canavezes et al. 1998; Hoyle et al. 2002; Hikage et al. 2002).
These investigations generally indicate consistency with the initial
Gaussianity.

For a statistical analysis based on the distribution of 
{\it luminous} objects,  however, non-Gaussianity produced
in the process of cosmic structure formation
should be properly considered.
Gravitational nonlinear evolution is one of the most important 
non-Gaussian sources. It has been well studied 
using a variety of models, including
a perturbative correction by the Edgeworth expansion (\cite{M1994}), 
direct numerical simulations (Matsubara, Suto 1996), an empirical 
lognormal model (Matsubara, Yokoyama 1996) and the Zel'dovich approximation
(\cite{Seto1997}).

Another non-Gaussian source is biasing, i.e., the statistical uncertainty 
between the spatial distribution of luminous objects and that of dark matter. 
Biasing is generally described by a nonlinear and 
stochastic function of the underlying mass density from various recent work 
(Dekel, Lahav 1999; \cite{BCOS1999, SLSDKW2001, TMJS2001}). 
The difficulty in a general prediction of a biasing expression comes from the fact 
that the biasing properties are sensitive to the complicated formation mechanism of
luminous objects. On the other hand, the biasing for dark matter halos, 
which are the most likely sites for galaxies and clusters,
is relatively easy to be modeled and thus well investigated on the basis of both
analytical and numerical approaches (Mo, White 1996; Taruya, Suto 2000;
\cite{YTJS2001}).  
In particular, dark halos with cluster-scale mass are usually supposed to 
have a fairly good one-to-one correspondence with galaxy clusters, at least
in a statistical sense. Therefore, biasing models for dark halos can be applied 
to upcoming galaxy cluster catalogs with an empirical relation between the halo mass 
and the temperature/luminosity of galaxy clusters (e.g., Kitayama, Suto 1997).
In this spirit, \citet{HTS2001} derives an analytical model 
of genus statistics for galaxy clusters with a nonlinear and stochastic halo
biasing model of \citet{TS2000}. 

A complementary approach to halo biasing is a direct
analysis of cosmological numerical simulations. The statistical
significance of previous studies, however, has been limited by the
number of particles and the box-size employed in the simulations. 
Since the mean separation of rich clusters of galaxies is 
$\sim 50h^{-1}$Mpc, a typical cosmological simulation in a $300h^{-1}$Mpc 
box has merely $\sim 200$ clusters. With this number of clusters, 
it is almost impossible to detect the cluster-mass dependence of various
statistical measures in a reliable manner (e.g., \cite{WMS1994}).
This situation may be significantly improved by using 
the Hubble Volume Simulation \citep{J2001}. Specifically,
we use the $\Lambda$CDM (Lambda Cold Dark Matter) data ($\Lambda$HVS,
hereafter) at $z=0$, which employ $N=10^9$ particles in a box-size of
$3h^{-1}$Gpc. The data enable us to quantitatively address the mass and
scale dependence of the halo biasing and the related statistics.

In the present work, we perform a numerical analysis 
of the genus for dark halos with the Hubble volume simulation in order to
clarify the uncertainty of the halo biasing in detecting
the initial non-Gaussianity.  
From a detailed comparison between the genus for dark halos and 
that for dark matter, we find that the halo biasing effect 
on the non-Gaussianity of genus is comparable to the non-Gaussianity
induced by the nonlinear gravitational evolution.
In order to characterize the halo biasing effect on the genus statistics,
we apply a perturbative formula developed by \citet{M1994},
and compare it with the simulated genus.
We find that the perturbative formula properly describes the halo genus 
at a smoothing scale larger than $50h^{-1}$Mpc. Also,  
the biasing effect on the genus for halos 
is shown to be well-approximated by the nonlinear
deterministic biasing up to the second-order terms 
in the mass density fluctuation.

The rest of the paper is organized as follows.  We briefly describe
the simulation data that we use and the halo catalogue in section
\ref{sec:simulation}.  We then discuss in section \ref{sec:jointpdf}
the biasing properties of dark halos while paying particular attention
to their nonlinear and stochastic nature.  We present a detailed study
of the topology of the spatial distribution of dark halos using the
genus statistics based on numerical simulations and a perturbative
analysis in section \ref{sec:genus}.  Finally section
\ref{sec:conclusion} is devoted to conclusions.

\section{Dark Matter halos in the Hubble Volume Simulation 
\label{sec:simulation}}

In the present work, we use the $z=0$ snapshot data of $\Lambda$CDM. 
The model assumes the density parameter of $\Omega_0=0.3$,
cosmological constant of $\lambda_0=0.7$, and the Hubble
constant in units of 100km s${}^{-1}$Mpc${}^{-1}$, $h=0.7$.  
The fluctuation amplitude,
smoothed over a top-hat radius of $8h^{-1}$Mpc, is set as
$\sigma_8=0.9$ so that the model satisfies both the cluster abundance
and the COBE normalization (e.g., \cite{ECF1996}, Kitayama, Suto 1997). 
The initial density field is computed with CMBFAST 
assuming baryon density parameter of $\Omega_{\rm b}h^2=0.0196$, 
and the random-Gaussian statistics \citep{C2000}.  The simulation box-size 
is $3000h^{-1}$Mpc and the softening length for the
gravitational force is set to be $100h^{-1}$kpc.  An initial
distribution of $N=10^9$ dark matter particles is generated using the
Zel'dovich approximation, and is advanced with the
Particle-Particle-Particle-Mesh (${\rm P^3M}$) code.  The mass of each
dark matter particle is set at $2.25\times 10^{12}h^{-1}\MO$.  \citet{HYSE2001}
made sure that the two-point clustering feature at $z=0$
in the $\Lambda$HVS is reliable even at an order of magnitude below the
mean particle separation $3h^{-1}$Mpc (still a few times larger than the
adopted softening length). Thus, the dynamic range of the simulation data
amounts to $\sim 10^4$ in length.

Dark halos are identified with a friends-of-friends finder with a
linking length of $b=0.164$ in units of the mean-particle separation.  We
define linked groups with the minimum number of member particles to be 30 
as dark halos in the current analysis. 
We divide those halos (about $1.5\times10^6$ in total) in five
different subgroups according to their mass (table~\ref{tab:halolist})
so that each subgroup includes about $3\times 10^5$ halos.  Because of
the huge size of the $\Lambda$HVS, the number of halos in each subgroup
is sufficiently large to characterize the mass dependence of various
statistical measures in a reliable manner. In the analysis discussed below, we
evaluate the error for each statistical measure from the
sample-to-sample variation among the results for eight subboxes
(box-size of $1500h^{-1}$Mpc) constructed out of the entire simulation box.
\begin{table}[thb]
  \caption{List of five subgroups of dark matter halos 
   in the $\Lambda$HVS.  \label{tab:halolist}}
 \begin{center}
 \begin{tabular}{ccc}
   \hline\hline
   subgroup & mass range [$h^{-1}M_\odot$] & number of halos \\ \hline
   SS & $6.7\times 10^{13}$ -- $8.0\times 10^{13}$ & 328449 \\ 
    S & $8.0\times 10^{13}$ -- $1.0\times 10^{14}$ & 321732 \\ 
    M & $1.0\times 10^{14}$ -- $1.3\times 10^{14}$ & 295204 \\ 
    L & $1.3\times 10^{14}$ -- $2.0\times 10^{14}$ & 300828 \\ 
   LL & $2.0\times 10^{14}$ -- $6.0\times 10^{15}$ & 314782 \\ \hline
    \end{tabular}
  \end{center}
\end{table}

\section{The Nature of Biasing of Dark Halos \label{sec:jointpdf}}

\begin{figure}[thp]
\begin{center}
\FigureFile(80mm,80mm){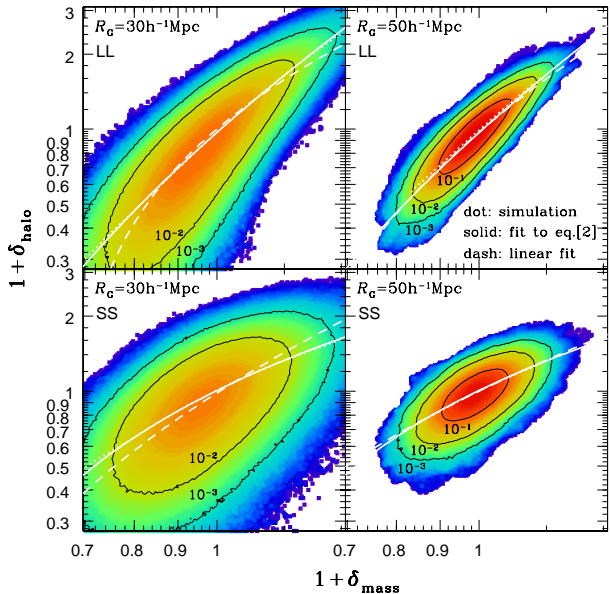}
\end{center}
\caption{Joint probability distribution functions,
 $P(\deltam, \deltah)$, of the halo and the dark
 matter density fields computed from the $\Lambda$HVS data with Gaussian
 smoothing; {\it Upper-left:} LL halos with $R_{\rm G}=30h^{-1}$Mpc,
 {\it Upper-right:} LL halos with $R_{\rm G}=50h^{-1}$Mpc, {\it
 Lower-left:} SS halos with $R_{\rm G}=30h^{-1}$Mpc, {\it Lower-right:}
 SS halos with $R_{\rm G}=50h^{-1}$Mpc. Contour lines show
 $P(\deltam,\deltah)=10^{-1}$, $10^{-2}$ and $10^{-3}$. 
 Dotted, dashed and solid lines 
 indicate the mean biasing [equation (\ref{eq:meanbias})] and the linear and
 the quadratic fits [equation (\ref{eq:fitbias})] 
 to the mean biasing respectively.}
 \label{fig:jointPDF_G}
\end{figure}

First we consider the nonlinear and stochastic nature of halo
biasing with particular attention paid to its mass- and scale-dependence.  For
this purpose, we compute the joint probability distribution function
(PDF) of the number density field of halos, $\deltah$, and the mass
density field of dark matter particles, $\deltam$.  Those density fields
are first evaluated at $128^3$ grids for each subbox data of particles
smoothed over a Gaussian radius of $R_{\rm G}$.
Then, the averaged value of the joint PDF over eight subboxes
is computed. Figure \ref{fig:jointPDF_G} shows the averaged joint PDF
for specific halo subgroups, LL and SS, at smoothing scales of
$30$ and $50h^{-1}$Mpc.

The dotted lines represent the mean biasing, $\bardeltah(\deltam)$,
averaged over $\deltah$ at a fixed value of $\deltam$:
\begin{equation}
\label{eq:meanbias}
\bardeltah(\deltam)=\int\deltah~P(\deltah | \deltam)~d\deltah .
\end{equation}
Here, $P(\deltah | \deltam)$ represents
the conditional PDF of the halo fields at a given $\deltam$, 
which is equal to the joint PDF, $P(\deltam, \deltah)$, divided by
the one-point PDF of dark matter, $P(\deltam)$. 
We fit the mean biasing to the following quadratic nonlinear model:
\begin{equation}
\label{eq:fitbias}
f_{\rm bias,fit}(\deltam) = b_1\deltam + \frac{b_2}{2}(\deltam^2 - \sigmamm^2),
\end{equation}
where $\sigmamm\equiv\langle\deltam^2\rangle^{1/2}$ 
and $\langle\cdot\cdot\cdot\rangle$
denotes the average over $P(\deltam)$.
The fitted values for the linear and the second-order biasing
coefficients, $b_1$ and $b_2$, in each subgroup 
are listed in table \ref{tab:fitbias}.
While a linear fit (dashed lines) with $b_2=0$ in equation
(\ref{eq:fitbias}) fails to approximate
the simulation result, a quadratic fit (solid lines) is quite
acceptable for our interest range in the biasing of cluster-scale halos
\citep{HTS2001}. 

On the other hand, the stochasticity in the biasing
is significantly affected by a shot-noise term in these halo samples 
due to the small number density of halos.
For example, the number of halos within the Gaussian smoothing volume,
$(2\pi)^{3/2}R_{\rm G}^3$ is around $22$ at $R_{\rm G}=50h^{-1}$Mpc, 
and then the shot-noise term amounts to $0.21(=1/\sqrt{22})$, 
which is comparable to the r.m.s. variance listed in table \ref{tab:fitbias}.
It is, therefore, not easy to extract the information concerning the
{\it pure} stochasticity in the clustering biasing from our halo catalogs.
We therefore focus on the non-linear biasing effect and compare the genus for
dark halos with perturbative predictions.
The effect of the stochasticity of biasing on the halo genus
is discussed in appendix 1 using a perturbative analysis.

\begin{table*}[thb]
  \caption{Linear and second-order biasing coefficients. 
\label{tab:fitbias}}
  \begin{center}
    \begin{tabular}{cccccccccc}
      \hline\hline	
      \multicolumn{1}{c}{}&
      \multicolumn{3}{c}{30$h^{-1}$Mpc}&
      \multicolumn{3}{c}{50$h^{-1}$Mpc}&
      \multicolumn{3}{c}{100$h^{-1}$Mpc} \\ \cline{2-10}
      \multicolumn{1}{c}{\raisebox{1.5ex}{subgroup}}&
      \multicolumn{1}{c}{$b_1^\ast$}&
      \multicolumn{1}{c}{$b_2^\dagger$}&
      \multicolumn{1}{c}{$\sigma^\ddagger$}&
      \multicolumn{1}{c}{$b_1^\ast$}&
      \multicolumn{1}{c}{$b_2^\dagger$}&
      \multicolumn{1}{c}{$\sigma^\ddagger$}&
      \multicolumn{1}{c}{$b_1^\ast$}&
      \multicolumn{1}{c}{$b_2^\dagger$}&
      \multicolumn{1}{c}{$\sigma^\ddagger$} \\ \hline
SS &  $1.69$ & $-0.72(\pm 0.05)$ & $0.32$ & $1.66$ & $-0.69(\pm 0.13)$ & $0.16$ & $1.63$ & $1.50(\pm 1.18)$  & $0.06$ \\
S &  $1.76$ & $-0.60(\pm 0.04)$ & $0.33$ & $1.76$ & $-0.76(\pm 0.14)$ & $0.16$ & $1.72$ & $-0.24(\pm 1.18)$  & $0.06$\\
M &  $1.88$ & $-0.31(\pm 0.04)$ & $0.35$ &$1.87$ & $-0.53(\pm 0.15)$ & $0.17$ & $1.88$ & $-1.95(\pm 1.05)$ & $0.06$\\
L &  $2.16$ & $0.38(\pm 0.05)$ & $0.36$ & $2.12$ & $-0.27(\pm 0.16)$ & $0.18$ & $2.05$ & $-0.40(\pm 1.08)$  & $0.06$\\
LL &  $2.94$ & $4.37(\pm 0.03)$ & $0.42$ &$2.97$ & $4.48(\pm 0.11)$ & $0.21$ & $3.05$ & $1.59(\pm 0.96)$  & $0.07$\\ \hline
    \end{tabular}
    \\

       ${}^\ast$ Linear biasing coefficients. \\
	${}^\dagger$ Second-order biasing coefficients with $1\sigma$ fitting error[equation (\ref{eq:fitbias})]. \\
	${}^\ddagger$ the amplitude of the rms density fluctuation of halos 
$\sigma\equiv\langle\deltah^2\rangle^{1/2}$ with Gaussian smoothing.
	\end{center}
\end{table*}
\section{Genus Statistics for Dark Halos in $\Lambda$HVS \label{sec:genus}}
\subsection{Definition and Computing Method of Genus}
Genus, $g(\nus)$, is defined as $-1/2$ times the Euler characteristic of
the isodensity contour of the density field $\delta$ 
at the threshold level of $\nus$ times the r.m.s. fluctuation $\sigma$. 
In practice, this is equal to (number of holes) $-$ 
(number of isolated regions) of the isodensity surface.
In the random-Gaussian field, genus per unit volume is given by 
the following analytical form:
\begin{equation}
\label{eq:genus_rd}
g_\RG(\nus) = \frac{1}{(2\pi)^2}\left(\frac{\sigma_1^2}{3\sigma^2}\right)^{3/2}
(1-\nus^2)\exp\left(-\frac{\nus^2}{2}
\right) ,
\end{equation}
where $\sigma_1 \equiv \langle|\nabla\delta|^2\rangle^{1/2}$ and
$\sigma \equiv \langle\delta^2\rangle^{1/2}$ 
($\nabla\delta$ denotes the spatial derivative of the density field 
$\delta$ and $\langle\cdot\cdot\cdot\rangle$
denotes the average over of the PDF of $\delta$ and $\nabla\delta$
(cf., \cite{D1970, A1981, BBKS1986, HGW1986}).  

\begin{figure*}
\begin{center}
\FigureFile(154mm,77mm){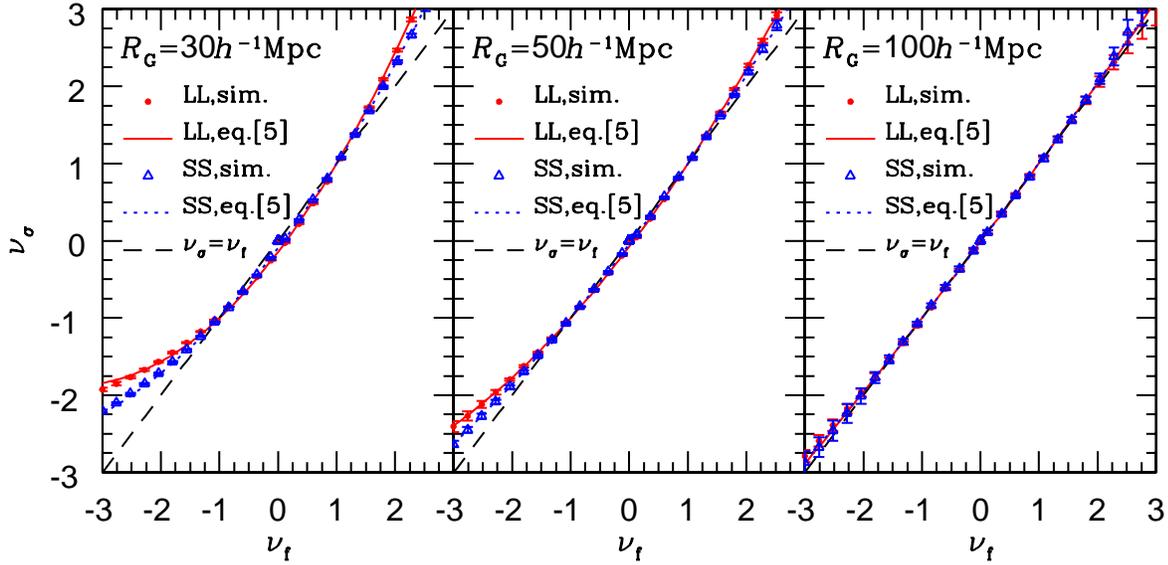}
\end{center}
\caption{Comparison between $\nuf$ and $\nus$ directly computed from
the density fields of LL (filled circles) and SS (open triangles) halo 
subgroups with sample-to-sample variations.
The perturbative expression up to the first order of $\sigma$ 
[equation (\ref{eq:nu_fvss_pb})] is also plotted for the LL subgroup (solid lines) and
the SS subgroup(dotted lines). From left to right, 
Gaussian smoothing scale is $30, 50,$ and $100h^{-1}$Mpc, respectively.
The dashed lines indicate the linear relation.
\label{fig:nu_fvss}}
\end{figure*}

In contrast, the genus is often expressed as a function of $\nuf$ instead of $\nus$
(e.g., Gott et al. 1989). The threshold $\nuf$ is used to 
parameterize the fraction $f$ of the volume 
that lies on the high-density side of the contour,
\begin{equation}
\label{eq:nuf}
f=\frac{1}{\sqrt{2\pi}}\int_{\nuf}^{\infty} e^{-x^2/2}dx.
\end{equation}
Figure \ref{fig:nu_fvss} shows a comparison between $\nus$ and $\nuf$.
The deviation from a linear relation represents the non-Gaussianity  
of the one-point PDF in the halo field smoothed with scales of
$30,~50$ and $100h^{-1}$Mpc from the left to right panels.
While the genus with $R_{\rm G}=100h^{-1}$Mpc approaches the random-Gaussian 
prediction, the non-Gaussianity of the one-point PDF is detectable 
with a smaller $R_{\rm G}$ and a larger mass of halo.
Up to the first order of perturbative correction in $\sigma$, 
the relation between $\nus$ and $\nuf$ is given by (\cite{M2003})
\begin{equation}
\label{eq:nu_fvss_pb}
\nus=\nuf+\frac{S^{(0)}}{6}(\nuf^2-1)\sigma,
\end{equation}
where we directly compute a skewness parameter, 
$S^{(0)}(\equiv\langle\nus^3\rangle/\sigma)$, from the halo field.
Figure \ref{fig:nu_fvss} shows that the lowest order approximation describes 
the simulated results very accurately at scales of our interest. 

If the evolved density field has an exact one-to-one correspondence with the
initial random-Gaussian field, then this transformation from $\nus$ to $\nuf$
removes the effect of evolution of the one-point PDF of the density field.
Under this assumption, the genus as a function of the volume fraction,
expressed as $g(\nuf)$, is sensitive only to the topology of the
isodensity contours, rather than evolution with time of the density
threshold assigned to a contour.  The limitations of the 
one-to-one mapping between the initial and evolved density fields
are critically examined by Kayo et al.  (2001).
In the following section, we present the results of the genus in 
terms of both $\nus$ and $\nuf$.

The computation of genus proceeds as follows: we first assign densities
of halos and dark matter particles on $128^3$ grids using the
cloud-in-cell method. The distance between each grid is sufficiently small
so as not to affect the density field smoothed over a scale of $30h^{-1}$Mpc.
The density fields are Fourier-transformed,
multiplied by the Gaussian  window with a smoothing length of 
$R_{\rm G}$, and then transformed back to a real
space. The {\it smoothed} density fields are used in defining
the isodensity surface with a given threshold, and the resulting genus is
evaluated with the CONTOUR 3D code \citep{weinberg88} by integrating the
deficit angles over the isodensity surface.

\subsection{Genus Statistics for Dark Matter and Dark Halos}

Since the one-point PDF of dark matter particles is empirically known to
be well approximated by the log-normal PDF (e.g. Coles, Jones 1991; 
Kofman et al. 1994; Kayo et al. 2001), it is natural to expect that it
is also the case for genus. In fact, \citet{MY1996} derived a genus
expression for dark matter assuming the nonlinear density field of dark
matter has a one-to-one mapping to its primordial Gaussian field. If one
adopts the log-normal mapping, their result can be explicitly written as
\begin{eqnarray}
\label{eq:genus_dm}
g_{\rm \LN}(\nus)&=&
\gmaxLN [1-x_\LN^2(\nus)] 
\,\exp \left[-\frac{x_\LN^2(\nus)}{2}\right] , \\
x_\LN(\nus) &\equiv& \frac{\ln[(1+\nus\sigma)\sqrt{1+\sigma^2}]}
{\sqrt{\ln(1+\sigma^2)}} , 
\end{eqnarray}
where the maximum value, $\gmaxLN$ is defined by
\begin{equation}
\gmaxLN  = \frac{1}{(2\pi)^2}
\left[\frac{\sigma_{\rm 1}^2}
{3(1+\sigma^2)\ln(1+\sigma^2)} \right]^{3/2} .
\end{equation}
Note that $g_{\rm \LN}$ in terms of $\nuf$ just reduces to the 
Gaussian prediction (\ref{eq:genus_rd}).
In reality, this one-to-one mapping assumption is much stronger than the
statement that the one-point PDF is empirically well approximated by
the log-normal PDF.

\begin{figure*}
  \begin{center}
    \FigureFile(80mm,80mm){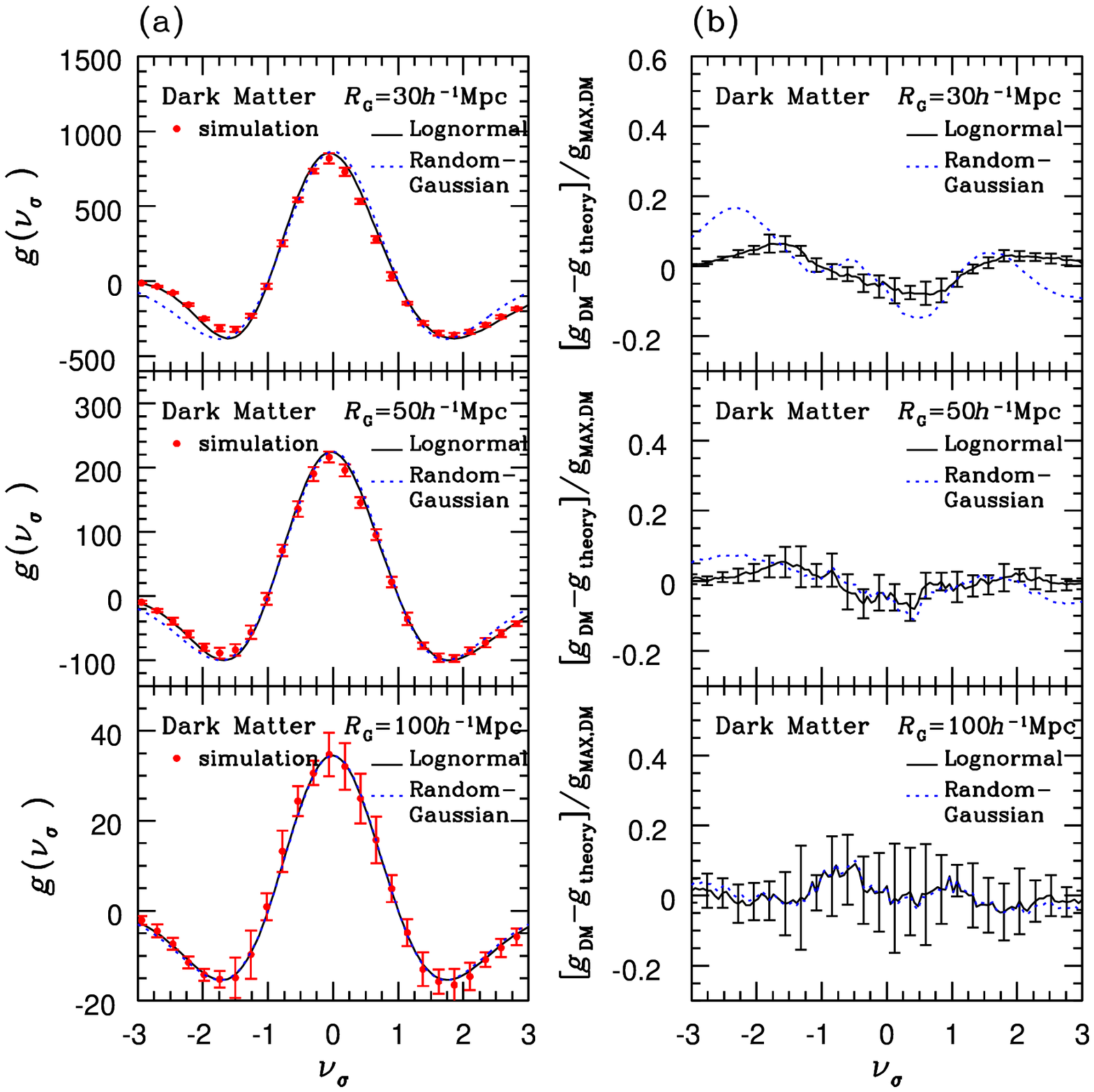}
    \FigureFile(80mm,80mm){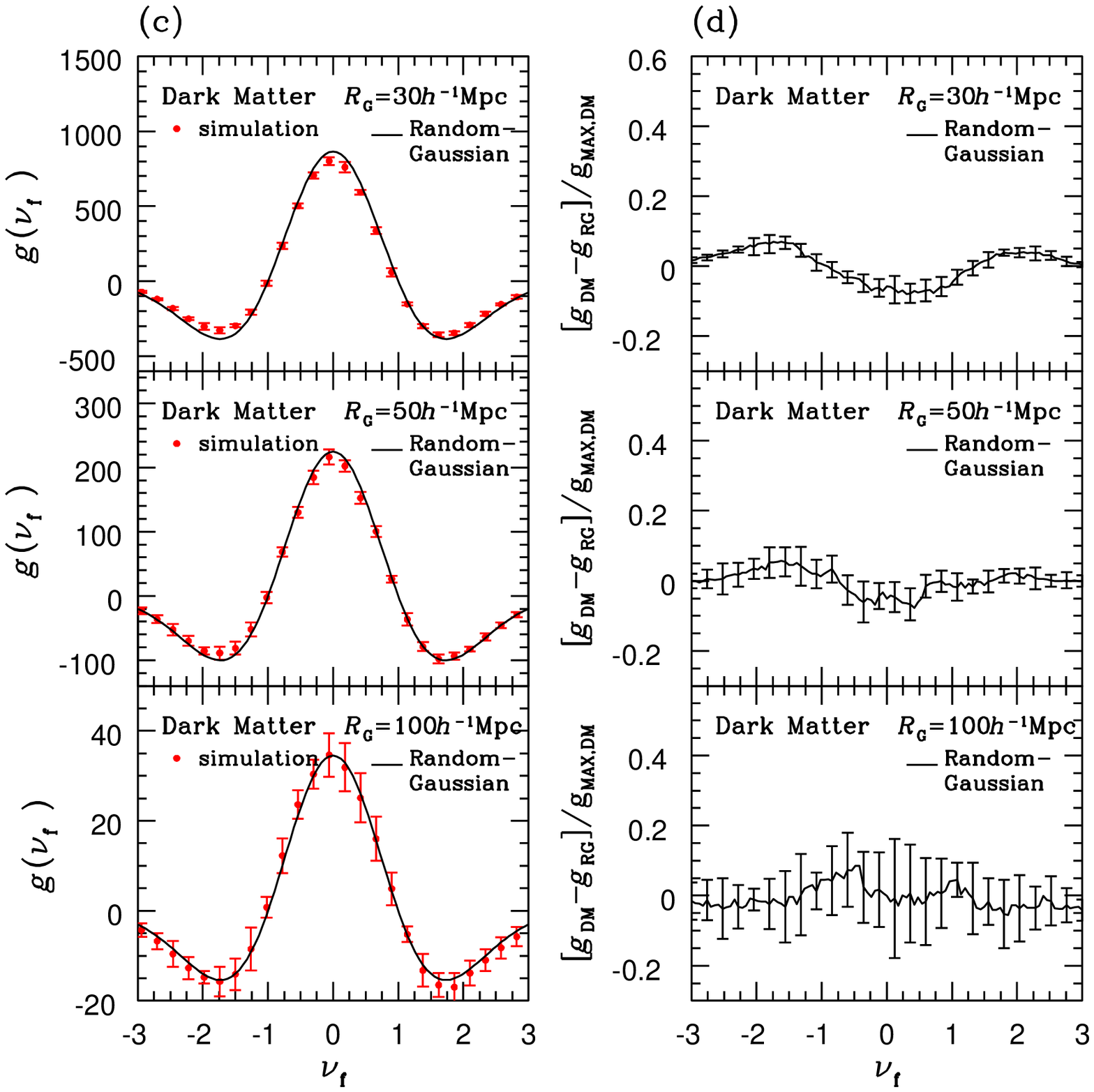}
  \end{center}
\caption{Panel (a): Genus 
for dark matter computed from $\Lambda$HVS with filled circles 
and the analytical genus expression in random-Gaussian 
with dotted lines [equation (\ref{eq:genus_rd})] and 
in log-normal statistics with solid lines [equation (\ref{eq:genus_dm})]. 
The top, middle, and bottom panels, respectively, indicate the results for smoothing
lengths of $30h^{-1}$Mpc, 
$50h^{-1}$Mpc, and $100h^{-1}$Mpc. 
This order of the smoothing scale is the same in the other panels.
Panel (b):
The difference between our simulation results (denoted by $g_{\rm\scriptscriptstyle DM}$) 
and the theoretical predictions (denoted by $g_{\rm\scriptscriptstyle 
theory}$; the random-Gaussian expression with dotted lines, 
the log-normal model with solid lines containing error-bars)
normalized by the amplitude of the simulated genus (denoted by 
$g_{\rm\scriptscriptstyle MAX,DM}$). 
Panels (c) and (d): Same as panels (a) and (b), but for
genus as a function of $\nuf$ [equation (\ref{eq:nuf})].
The simulated results are compared with
the random-Gaussian expression with error-bars (denoted by $g_{\rm RG})$}.
\label{fig:genus_mass_G}
\end{figure*}

We compare the genus for dark matter averaged over eight sub-boxes
with the analytical expressions [equations (\ref{eq:genus_rd}) and
(\ref{eq:genus_dm})] multiplied by the volume of the sub-boxes,
$(1500h^{-1}{\rm Mpc})^3$. The calculations of $\sigma$ and
$\sigma_{\rm 1}$ in the analytical predictions are based on the power
spectrum by Peacock and Dodds (1996).  Figure \ref{fig:genus_mass_G}
shows that genus as a function of $\nus$ [panel (a)], the difference,
between the genus for the simulated dark matter $g_{\rm DM}$ and the
analytical predictions $g_{\rm theory}$, normalized by the maximum
value of $g_{\rm DM}$, $g_{\rm\scriptscriptstyle MAX,DM}$ [panel (b)],
and the same as panels (a) and (b), but as a function of $\nuf$
[panels (c) and (d)].

\begin{figure*}
\begin{center}
\FigureFile(80mm,80mm){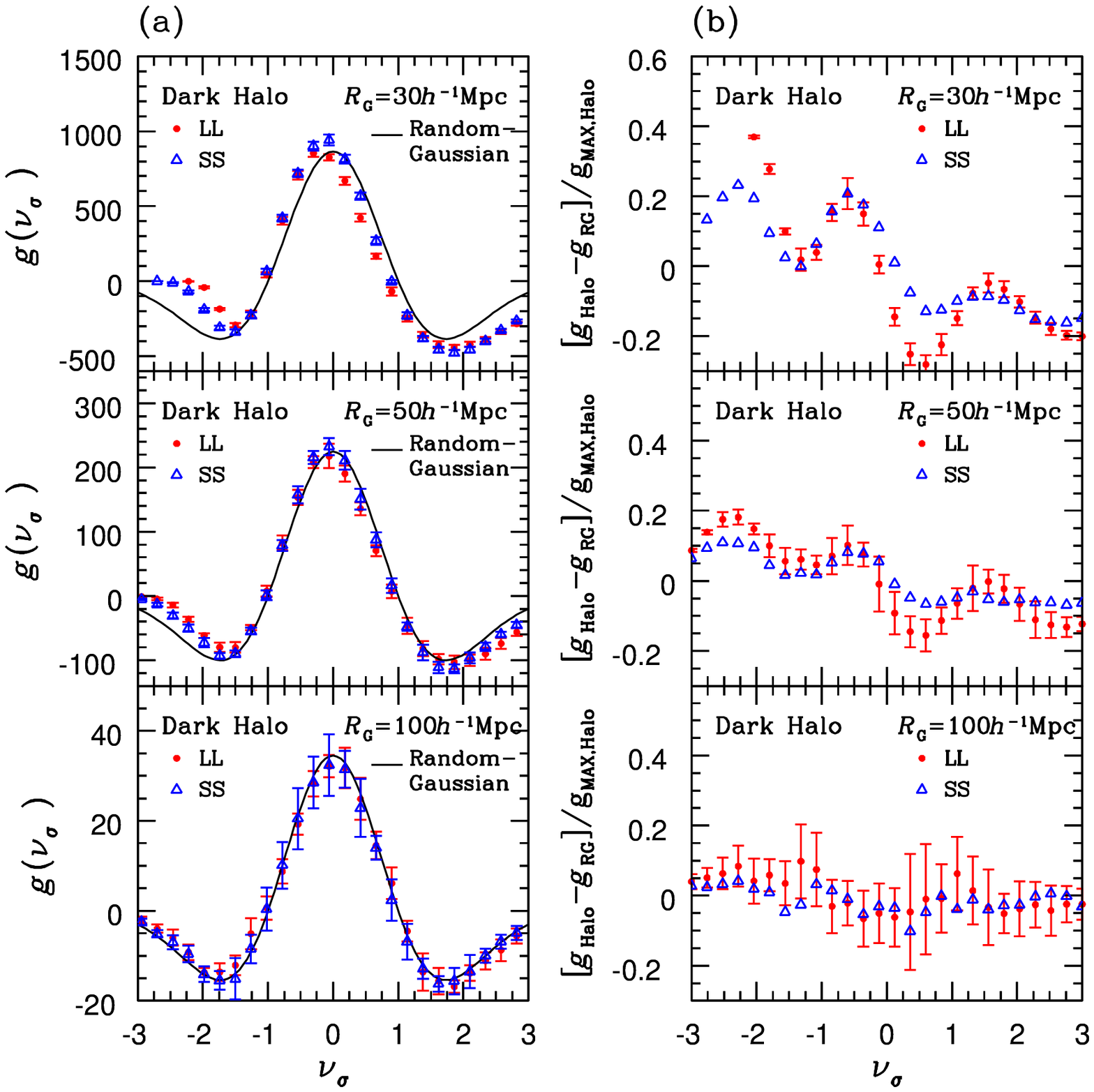}
\FigureFile(80mm,80mm){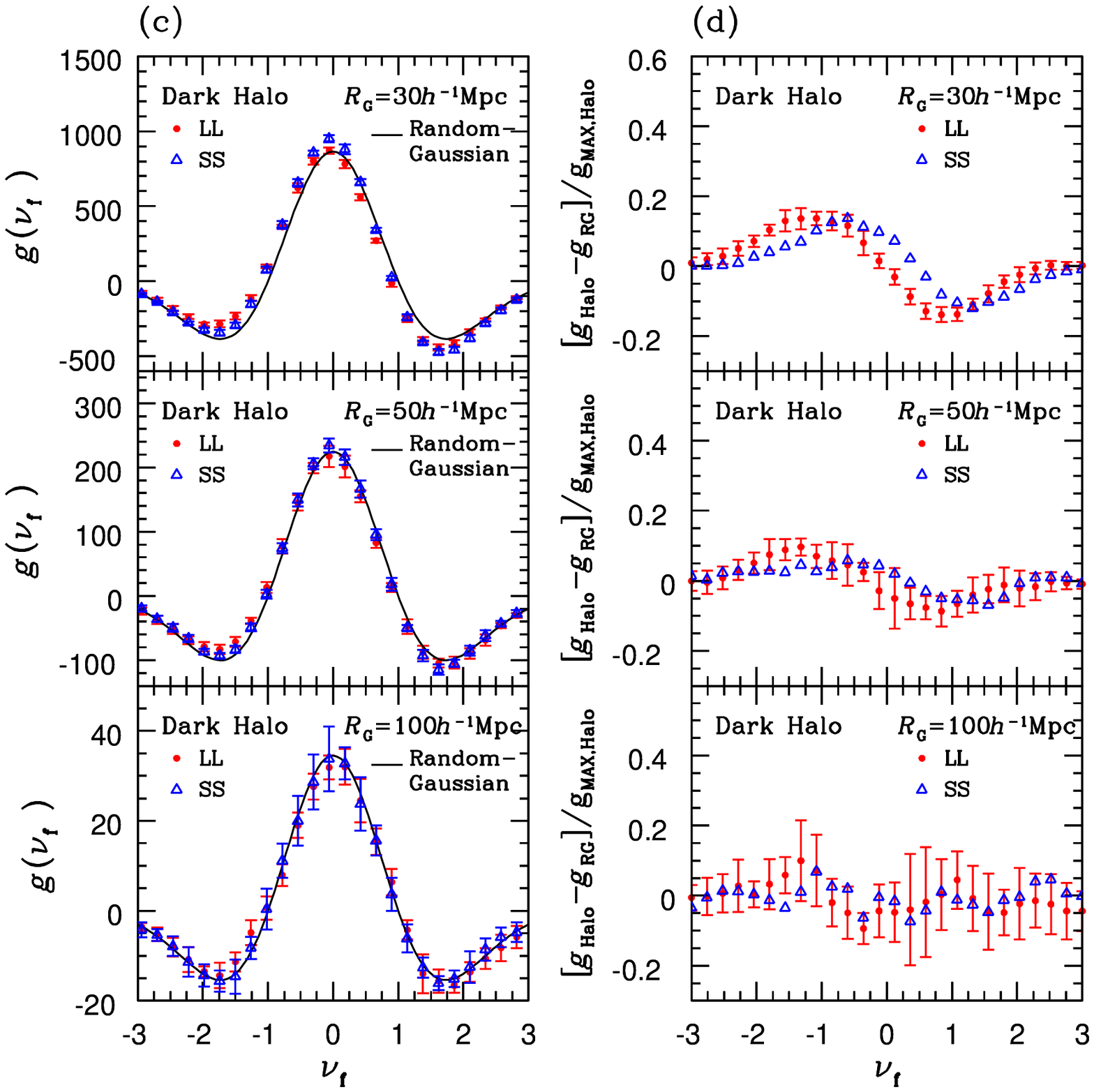}
\caption{Same as figure \ref{fig:genus_mass_G} but for dark halos 
in LL (filled circles) and SS (open triangles) subgroups. 
The sample-to-sample variance of genus is plotted with error-bars 
only for the LL subgroup.
Panels (a) and (b): Comparison between the genus for halos (denoted by 
$g_{\rm\scriptscriptstyle Halo}$) 
with the simulated genus for dark matter (denoted by 
$g_{\rm\scriptscriptstyle DM}$)
as a function of $\nus$. Panels (c) and (d): 
Comparison between genus for halos 
as a function of $\nuf$ with equation (\ref{eq:genus_rd}).
\label{fig:genus_halo_G}}
\end{center}
\end{figure*}
We estimate the sample-to-sample variance of the simulated genus 
$\Delta g(\nu)$ by
\begin{equation}
\Delta g(\nu)=\left\{\sum_{i=1}^{8}[g_i(\nu)-\bar{g}(\nu)]^2/7\right\}^{1/2},
\end{equation}
where $g_i(\nu)$ denotes the value of the genus on the contour surface 
at the threshold $\nu$ estimated from sub-box $i$;
$\bar{g}(\nu)$ is the value of the genus averaged over eight sub-boxes.

Panels (a) and (b) show that the empirical prediction (\ref{eq:genus_dm}) 
well explains our current results for dark matter including 
both the amplitude and the shape of the genus  in our interested scale.
This is consistent with the numerical results by Matsubara and Yokoyama (1996).Panels (c) and (d) show that the genus for dark matter as a function of $\nuf$ 
agrees with the analytical prediction for a Gaussian random field
[equation (\ref{eq:genus_rd})].

\begin{figure}
\begin{center}
\FigureFile(80mm,80mm){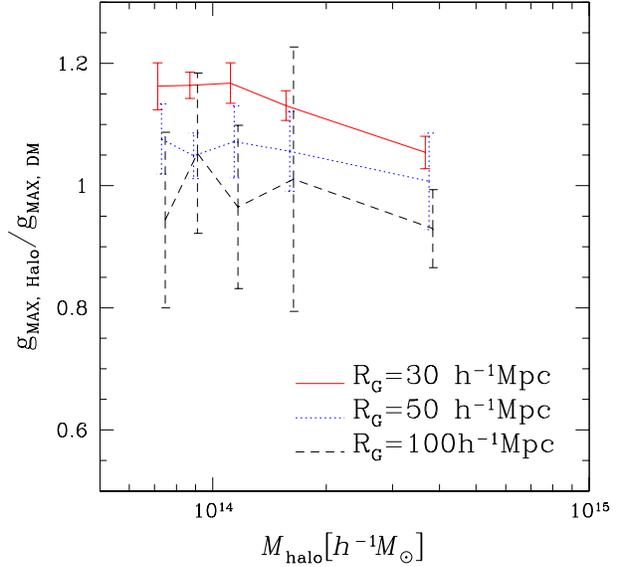}
\caption{Ratio of the amplitude of genus for halos, 
$g_{\rm\scriptscriptstyle MAX,Halo}$, to that for dark matter,
$g_{\rm\scriptscriptstyle MAX,DM}$ as a function of the mean halo mass 
for each subgroup. The smoothing scale $R_{\rm G}$ is $30$ (solid), 
$50$ (dotted), and $100 h^{-1}$Mpc (dashed).
\label{fig:genus_amplitude}}
\end{center}
\end{figure}

We then plot the genus for dark halos of two specific subgroups, LL and SS
in figure \ref{fig:genus_halo_G}. While panels (a) and (b) show 
comparisons of the halo genus with the simulated genus for mass 
as a function of $\nus$, the panels (c) and (d) represent the results
in terms of $\nuf$.
The differences in the genus for halos from the simulated genus for dark matter
[panel (b)] or the random-Gaussian prediction [panel (d)] are normalized 
by the maximum value of the genus for halos, $g_{\rm\scriptscriptstyle MAX, Halo}$.
Figure \ref{fig:genus_halo_G} shows that the non-Gaussianity induced by
halo biasing is comparable to that by the non-linear gravitational
evolution, and that the shape and the amplitude of the halo genus are 
almost insensitive to the halo mass.
The result suggests that the influence of biasing can be corrected
perturbatively. To see the weak dependence of the genus on the halo mass
more clearly, we plot in figure \ref{fig:genus_amplitude} 
the ratio of the genus amplitude,  
$g_{\rm\scriptscriptstyle MAX,Halo}/g_{\rm\scriptscriptstyle MAX,DM}$,
i.e., the maximum value of the genus amplitude for dark halos divided by
that for dark matter.
The results for each halo subgroup are plotted as a function of the
averaged halo mass. While the increase of the smoothing scale significantly
changes the genus amplitude in figure \ref{fig:genus_halo_G},
the relative amplitude in figure \ref{fig:genus_amplitude} almost
remains unchanged for a wide range of the halo mass.
Furthermore, the amplitude of halo genus is very close to that of dark matter,
with only $6$ -- $16$ percent deviations at $R_{\rm G}=30h^{-1}$Mpc.
This is indeed consistent with the prediction of Hikage et al. (2001)
based on the one-to-one mapping assumption from the genus for dark matter.
Therefore, one can conclude that the biasing effect on halo genus on the
cluster-mass scale is generally small, and that the perturbative analysis
based on Edgeworth formula is safely applicable at $R_{\rm G}\ge 30h^{-1}$Mpc.

\subsection{Comparison with Perturbation Theory}
Having established the weak influence of the biasing, 
we wish to further examine the halo genus for a better
understanding of the nature of biasing. For this purpose,
we attempt to model the halo genus using
a perturbative approach with particular emphasis on the nonlinearity 
of the biasing effect.
We start with Matsubara's expression for the genus
density in a weakly non-Gaussian field (Matsubara 1994, 2003).  
In the lowest-order approximation, his result reduces to
\begin{eqnarray}
\label{eq:pt}
g_\PT(\nus)= &
-\frac{1}{(2\pi)^2}\left(\frac{\sigma_1^2}{3\sigma^2}\right)^\frac{3}{2}
\exp\left(-\frac{\nus^2}{2}\right) \left\{H_2(\nus)+ \right.\nonumber \\
\left[\frac{S^{(0)}}{6}\right. 
& \left.\left.H_5(\nus)+S^{(1)}H_3(\nus) +  S^{(2)}H_1(\nus)\right]\sigma\right\},
\end{eqnarray}
where $H_n(\nus)$ are the Hermite polynomials and $S^{(n)}$ represent the
skewness parameters, defined as
\begin{eqnarray}
\label{eq:skew}
S^{(0)}  =  \frac{\langle\nus^3\rangle}{\sigma}, 
\quad &
S^{(1)} = -\frac{3}{4}
\frac{\langle\nus^2(\nabla^2\nus)\rangle\sigma}{\sigma_1^2}, 
\nonumber \\
S^{(2)}  = & -\frac{9}{4}
\frac{\langle(\nabla\nus\cdot\nabla\nus)(\nabla^2\nus)\rangle\sigma^3}
{\sigma_1^4}.
\end{eqnarray}
With the relation between $\nus$ and $\nuf$ [equation (\ref{eq:nu_fvss_pb})],
the genus as a function of $\nuf$ is rewritten by
\begin{eqnarray}
\label{eq:pt_f}
g_\PT(\nuf)&=&
-\frac{1}{(2\pi)^2}\left(\frac{\sigma_1^2}{3\sigma^2}\right)^\frac{3}{2}
\exp\left(-\frac{\nuf^2}{2}\right) \{H_2(\nuf) +  \nonumber \\
  \left[(S^{(1)}\right. & - & \left.\left.S^{(0)})H_3(\nuf) 
+(S^{(2)}-S^{(0)})H_1(\nuf)\right]\sigma\right\}.
\end{eqnarray}
In the case of weak non-Gaussianity induced by the nonlinear
gravitational evolution, the above expression proves to be in good
agreement with the genus computed from the earlier N-body
simulation data (Matsubara, Suto 1996; Matsubara, Yokoyama 1996). 
We extend those previous studies
and examine whether or not equations (\ref{eq:pt}) and (\ref{eq:pt_f}) are 
also applicable to
the weak non-Gaussianity induced by the halo biasing (in addition to the
nonlinear gravitational evolution).

\begin{figure*}
\begin{center}
\FigureFile(80mm,80mm){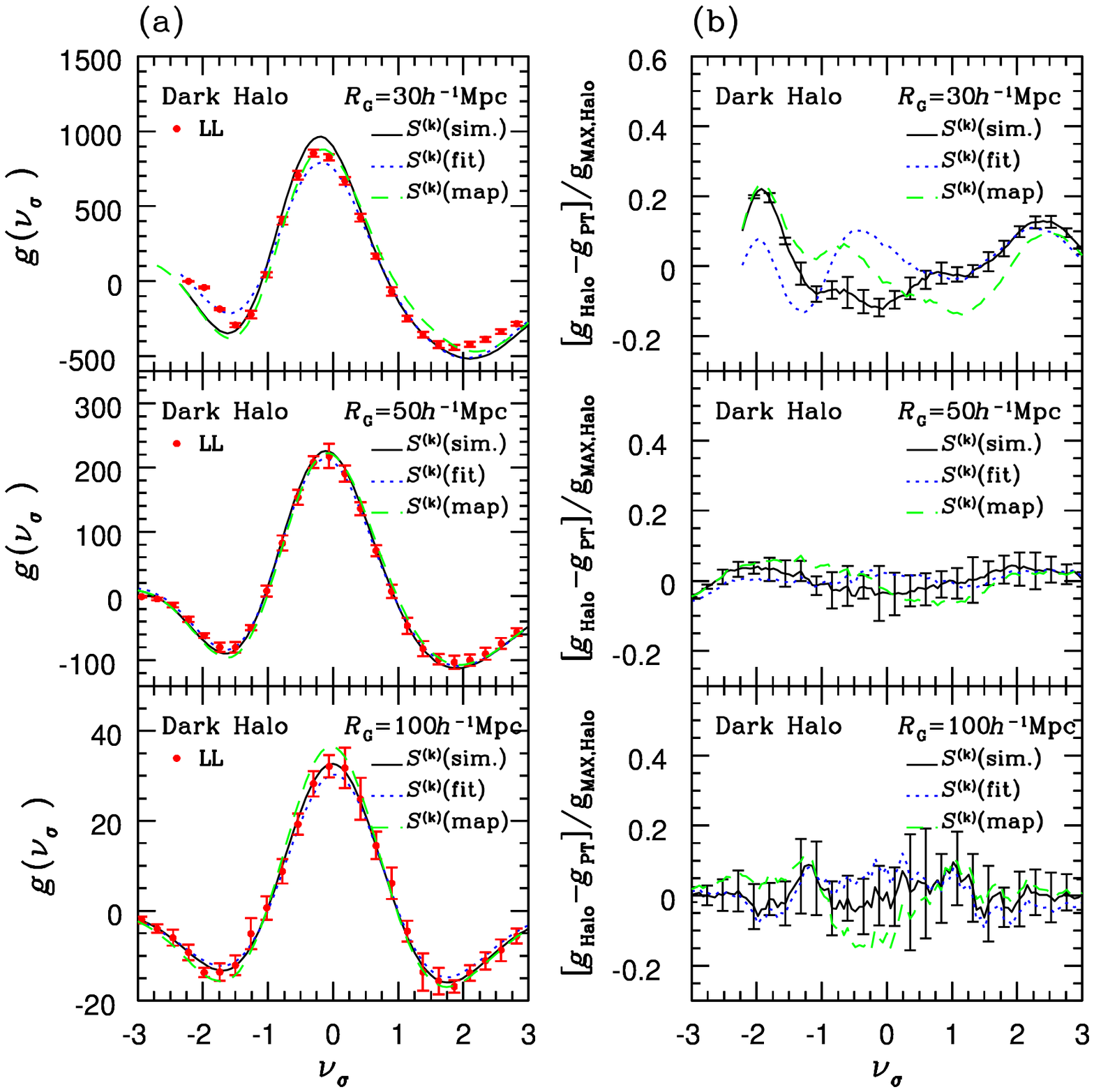}
\FigureFile(80mm,80mm){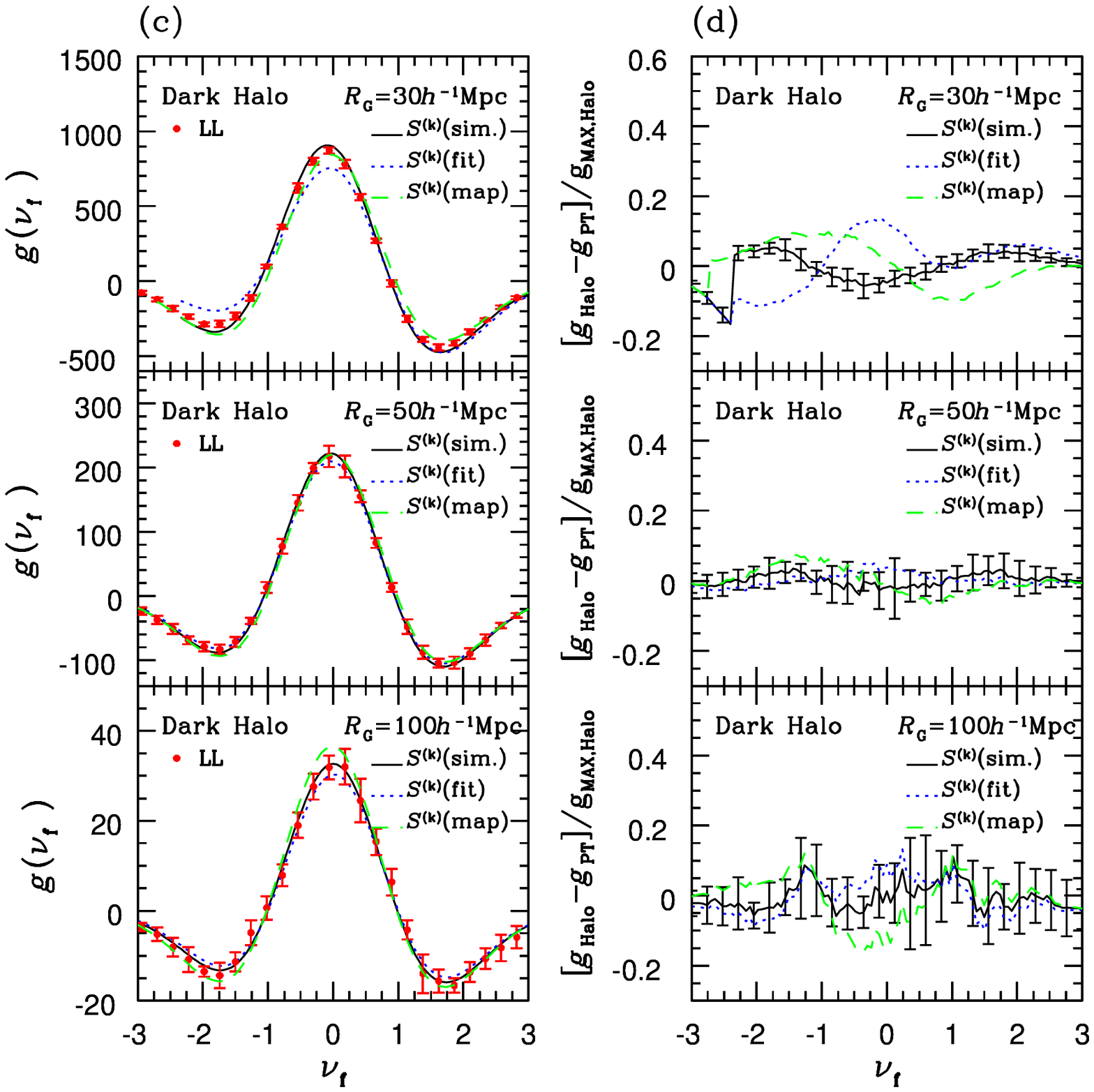}
\end{center}
\caption{
Same as figure \ref{fig:genus_halo_G}, but for a comparison of the simulated genus
 for LL halos (denoted by $g_{\rm Halo}$) with the second-order perturbative 
 formula [equation (\ref{eq:pt})] (denoted by $g_{\rm PT}$). 
 The skewness parameters in the perturbative
 formula are calculated in the following three ways: (i)
 a direct computation from the simulated halo density field
 (solid, labeled by `sim.'),
 (ii) fitting of the simulated halo genus 
 to the perturbative formula with skewness parameters put 
 as free parameters (dotted, labeled by `fit'), and (iii)
 direct computation from the $\bardeltah$ field 
 obtained by one-to-one mapping from the dark matter field with
 the simulated mean biasing (dashed, labeled by `map').
 The error-bars, which indicate the sample-to-sample variance [equation(9)], 
 are plotted only for the genus with skewness parameters calculated by (i).}
\label{fig:genus_perturb_GA}
\end{figure*}
Figure \ref{fig:genus_perturb_GA} represents a comparison between the
simulated results for the most strongly biased 
halo subgroup LL and the perturbative model of genus
as a function of $\nus$ in panels (a) and (b) and of $\nuf$ 
in panels (c) and (d).
We explicitly show the differences between the simulation results, $g_{\rm Halo}$, 
and the prediction of the perturbative model, $g_{\rm\scriptscriptstyle PT}$, 
divided by the amplitude of the simulated genus, $g_{\rm\scriptscriptstyle 
MAX, Halo}$, in panel (b) for $g(\nus)$ 
and panel (d) for $g(\nuf)$. 
For the prediction of the perturbative model 
in equations (\ref{eq:pt}) and (\ref{eq:pt_f}), the r.m.s. fluctuation of the
halo density field, $\sigma=\langle\deltah^2\rangle^{1/2}$, and 
that of its first spatial derivative field,
$\sigma_1=\langle|\nabla\deltah^2|\rangle^{1/2}$, 
are computed directly from the halo density field. We compute the 
skewness parameters $S^{(0)}$, $S^{(1)}$, and $S^{(2)}$ 
using three different methods:
(i) a direct computation from the halo density field (labeled by `sim.'), 
(ii) fitting of our simulated genus to the perturbative 
formula as free parameters (labeled by `fit'), and
(iii) a direct computation from the $\bardeltah$ field (labeled by `map'), 
based on one-to-one mapping [equation (\ref{eq:meanbias})],
assuming the nonlinear deterministic biasing.
For computations of (i) and (iii), 
the gradient and Laplacian values of the density fields 
in equation (\ref{eq:skew}) are evaluated by making the
corresponding spatial derivative in Fourier space, respectively.
As for (ii), the range of the fitting is restricted
from $\nus=$Max$(-1/\sigma,-2.5)$ to $\nus=2.5$.

Figure \ref{fig:genus_perturb_GA} shows that the perturbative formula
[equation (\ref{eq:pt})] is valid for the halo genus
at $R_{\rm G}\ge 50h^{-1}$Mpc. The approximation by the perturbative formula
is also good in $|\nu|\le 1.5$ at $R_{\rm G}=30h^{-1}$Mpc, 
although the large deviation in $|\nu|\ge 1.5$ 
might be attributed to the limitation of the Edgeworth expansion,
which is also seen in the dark matter case (Matsubara and Suto 1996).
The difficulty to distinguish the perturbative prediction using
the $\bardeltah$ field from other genus
implies that the biasing effect on the halo genus is well approximated by
nonlinear deterministic biasing.

\begin{figure*}
\begin{center}
\FigureFile(160mm,160mm){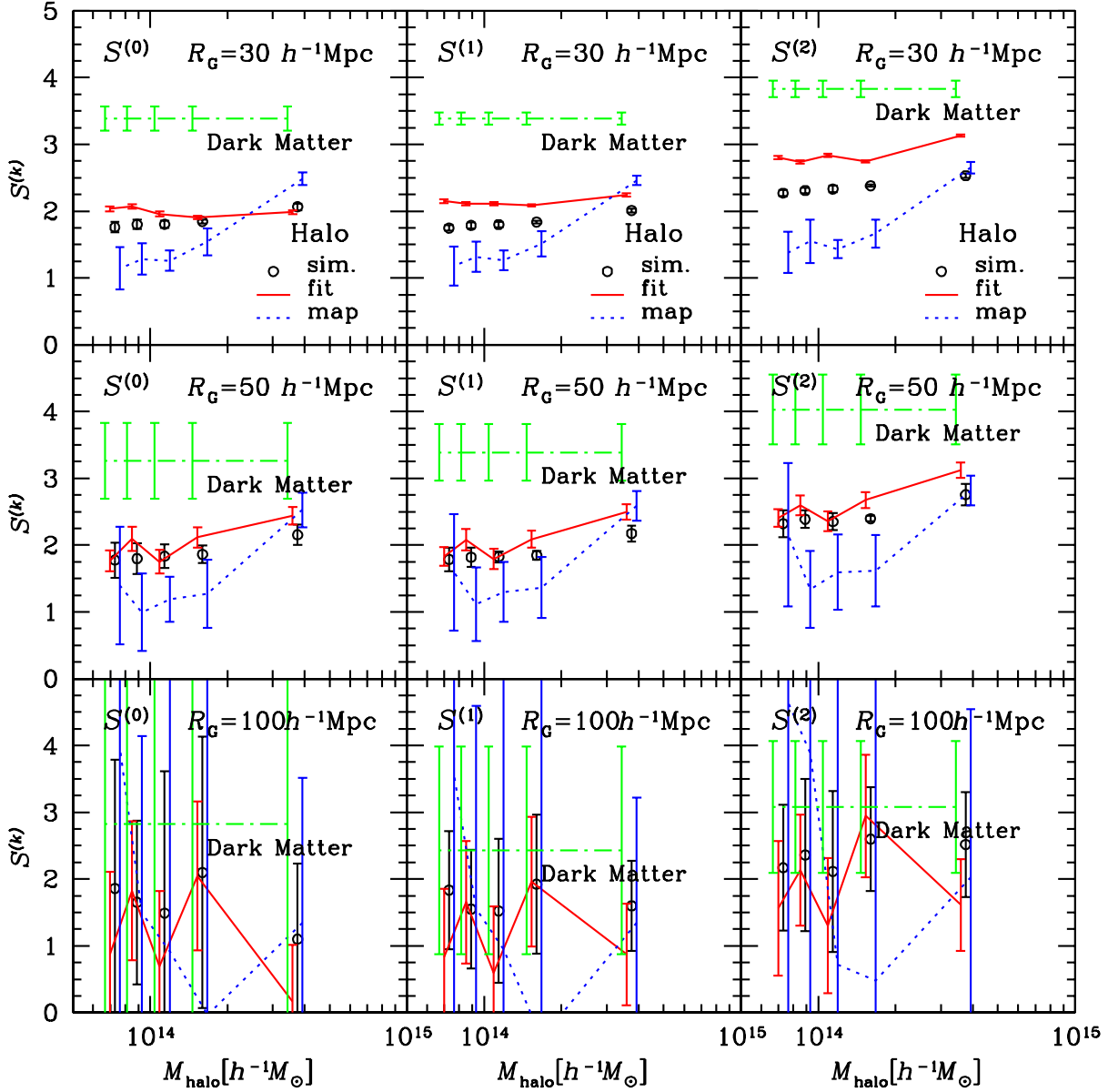}
\caption{Skewness parameters, $S^{(0)}$ ({\it left}), $S^{(1)}$ ({\it middle})
 and $S^{(2)}$ ({\it right}) as a function of halo mass averaged over
each subgroup with $R_{\rm G}=30h^{-1}$Mpc ({\it top}), 
$R_{\rm G}=50h^{-1}$Mpc ({\it middle}) 
and $R_{\rm G}=100h^{-1}$Mpc ({\it bottom}).
Labels of 'sim'., 'fit' and 'map' have the same meanings 
as figure \ref{fig:genus_perturb_GA}.
Dashed lines indicate skewness parameters for the dark matter field. 
The horizontal position of each plot is slightly staggered for easy view.}
\label{fig:skew_fit_G}
\end{center}
\end{figure*}

Next, we examine whether or not one can reproduce the
best-fit values for those skewness parameters from the independent
theoretical models and/or simulation data.
Figure \ref{fig:skew_fit_G} plots the skewness parameters 
calculated from the above three methods.
A weak dependence of the skewness parameters on the halo mass
is consistent with a model prediction by Mo et al. (1997).
Fitting method (ii) reproduces the skewness parameters well
at a smoothing scale of $50h^{-1}$Mpc, while the approximation
by the perturbative formula breaks down on the scale of $30h^{-1}$Mpc,
especially for the $S^{(2)}$ term.
It is almost impossible to reproduce skewness parameters
accurately at a smoothing scale of $100h^{-1}$Mpc, because
the statistical error is large due to the small value of the genus.

A good agreement of the skewness parameters from the $\bardeltah$ field
again suggests that the approximation 
by the nonlinear deterministic biasing works well in the halo genus
over the cluster-mass scales.
Recalling the fact that the nonlinearity in the halo biasing is well 
described by the linear and the quadratic terms shown in section 3,
an accurate prediction of the halo genus is feasible from
appropriate modeling of the two biasing coefficients.
The reason why the stochastic effect of the biasing 
is nearly negligible can be understood from the fact that
the stochastic components are almost cancelled out by each other
in the form of skewness parameters (see appendix 1).

\section{Conclusions \label{sec:conclusion}}

We evaluate the extent to which halo biasing affects 
the genus for dark halos of cluster-scale mass using the Hubble 
volume simulations. 
The huge volume of the data enables us to perform the most reliable systematic
analysis of the genus for dark halos and to calculate the sample-to-sample
error corresponding to the cosmic variance.
Through a detailed comparison of the genus for dark halos with that for the mass
distribution,
we find that the non-Gaussianity induced by the biasing is comparable to
that by the nonlinear gravitational evolution.
To characterize the biasing effect on the genus for dark halos,
we extensively apply a perturbative formula developed by Matsubara (1994).
We find that the perturbative formula is indeed very accurate for 
the genus of dark halos at a smoothing length of
$R_{\rm G} \ge 50h^{-1}$Mpc. The skewness parameters obtained from  
a fitting of the perturbative formula 
consistently reproduce the results directly
measured from the density field.
We also find that the nonlinear deterministic biasing well describes 
the biasing effect on the genus for dark halos.
Moreover, the nonlinearity of biasing can be
approximated by only the linear and quadratic terms. A prediction of genus
for the dark halo on the cluster-scale mass scale is thus possible 
from an appropriate modeling of these two parameters.

Our results show that the contribution of
biasing stochasticity to the genus statistics is
very small at $R\ge 30 h^{-1}$Mpc and that
the dependence of the halo mass is safely negligible within the 
sampling error-bars. On the other hand, both the 
non-linearity of the biasing and the gravitational evolution 
alter the shape of the genus, depending on the smoothing scale.
In a weakly nonlinear regime, one can even discriminate these two effects 
using perturbative theory. In light of this, 
our formulation, including the nonlinearity of biasing and
gravitational evolution in a previous paper (\cite{HTS2001}), 
may provide a theoretical guide for characterizing the real observational data.  
At the current observational level, however, 
the above biasing effect is very difficult to detect 
because the biasing effect on the genus 
is small even in a much larger volume size of $(1500h^{-1}{\rm Mpc})^3$,
as shown in figure \ref{fig:genus_halo_G}.
The other observational effects, including the redshift distortion and
the light-cone effect, are also small in the present observational status
(Matsubara, Suto 1996; Colley et al. 2000). 
Therefore, the initial Gaussianity 
can be probed directly by the genus from the upcoming galaxy cluster 
catalogs within the sampling errors.

\bigskip

We thank the referee, David Weinberg, for a critical reading and 
useful comments. We also thank Naoki Yoshida for providing 
the dark matter and halo density fields from the Hubble Volume simulation 
and for useful comments and Y. Ohta for
providing his program to make contour plots. 
Numerical computations were carried out 
at ADAC (the Astronomical Data Analysis Center) of the National 
Astronomical Observatory, Japan (project ID:mys02) 
and at KEK (High Energy Accelerator 
Research Organization, Japan). One of us (A.T.) acknowledges the support
by the Grant-in-Aid for Scientific Research for JSPS (No. 1470157).
This research was supported 
in part by the Grant-in-Aid by the Ministry of Education, Culture, 
Sports, Science and Technology of Japan (07CE2002, 12640231) to RESCEU, 
and by the Supercomputer Project (No.99-52, N0.00-63) of KEK.


\appendix

\section{Influence of Biasing Stochasticity in a Weakly Nonlinear Regime}

In this appendix, we discuss the influence of 
biasing stochasticity on the halo genus in a weakly nonlinear regime.
For this purpose, based on the perturbative formula by Matsubara (1994)
[equations (\ref{eq:pt}) and (\ref{eq:skew})], we evaluate the skewness 
parameters. Suppose that the halo density
field is split into the deterministic and the residual stochastic terms;
\begin{equation}
\label{eq:deltaheh}
\deltah = \bardeltah+\epsilonh ,
\end{equation}
where the mean deterministic term is formally defined by
\begin{equation}
\bardeltah(\deltam)=\int\deltah\,\, P(\deltah|\deltam)d\deltah , 
\end{equation}
in terms of the conditional PDF of the halo fields at a given 
$\deltam$.  Note that equation (\ref{eq:deltaheh}) implies that 
$\langle \epsilonh F(\deltam)
\rangle = 0$ for any function $F(\deltam)$ because of the relation
$\langle \deltah F(\deltam) \rangle = \langle \bardeltah F(\deltam) \rangle$.

In a weakly nonlinear regime of the dark matter density field, 
the nonlinearity 
in the biasing may be well approximated by the following quadratic form:
\begin{equation}
\label{eq:nonlinearbias}
\bardeltah(\deltam) = b_1\deltam + \frac{b_2}{2}(\deltam^2 - \sigmamm^2) + \cdots,
\end{equation}
with the quantity $\sigmamm^2$ being $\langle\deltam^2\rangle$.
In this case, the variance of the halo and the covariance of the halo 
and dark matter are evaluated as follows: 
\begin{eqnarray}
\label{eq:sigmahalo}
\sigmahh^2 &\equiv& \langle \deltah^2 \rangle
\approx b_1^2 \sigmamm^2 + \langle \epsilonh^2 \rangle, \\
\sigmahm^2 &\equiv& \langle \deltah \deltam \rangle 
\approx b_1 \sigmamm^2 .
\end{eqnarray}
The above results can be translated to the skewness of the halo
density field as
\begin{eqnarray}
\label{eq:deltah3}
\langle \deltah^3 \rangle &\approx & b_1^3 \sigmamm^4 
\left( S_{\rm m}^{(0)} + 3 \frac{b_2}{b_1}\right. \nonumber \\  
&+ & \left.\frac{3 b_1 \langle \epsilonh^2 \deltam \rangle + \langle \epsilonh^3 \rangle}
{b_1^3 \sigmamm^4} \right),  
\end{eqnarray}
where the quantity $S_{\rm m}^{(0)}$ means the skewness of 
the mass density field, 
$S_{\rm m}^{(0)}\equiv \langle \deltam^3 \rangle/ \sigmamm^4$. 
Then, one obtains an expression for the normalized skewness, $S_{\rm
h}^{(0)}$, for the halo density field as
\begin{eqnarray}
S_{\rm h}^{(0)} &\equiv&
 \frac{\langle\deltah^3\rangle}{\sigmahh^4} \approx  \frac{1}{b_1}
\left( S_{\rm m}^{(0)} + 3 \frac{b_2}{b_1}\right.  \nonumber \\ 
& + & 
\left.\frac{3 b_1 \langle \epsilonh^2 \deltam \rangle + \langle \epsilonh^3 \rangle}
{b_1^3 \sigmamm^4} \right) 
\left( 1 + \epsilon_{\rm scatt}^2 \right)^{-2}, 
\label{eq: skewness_h_0}
\end{eqnarray}
where we adopt a measure for the stochasticity of biasing, 
$\epsilon_{\rm scatt}$ (Dekel, Lahav 1999; Taruya, Suto 2000),
\begin{equation}
\epsilon_{\rm scatt}^2 \equiv
\frac{\sigmamm^2 \langle \epsilonh^2 \rangle }{\sigmahm^4}
\approx 
\frac{1}{b_1^2} \frac{\langle \epsilonh^2 \rangle}{\sigmamm^2}. 
\end{equation}
In deriving the above results, we did not treat the stochastic 
term, $\epsilonh$,  as a small quantity, 
which might not be guaranteed even in a weakly nonlinear regime.  

Equation (\ref{eq: skewness_h_0}) is a natural extension of the frequently 
used expression within the framework of deterministic nonlinear biasing 
(e.g., Fry, Gazta\~naga 1993);  
\begin{eqnarray}
S_{\rm h}^{(0)} =
\frac{1}{b_1}
\left( S_{\rm m}^{(0)} + 3 \frac{b_2}{b_1} \right),   
\label{eq: skewness_FG1993}
\end{eqnarray}
which is only valid in absence of stochasticity, $\epsilonh=0$. 
Compared (\ref{eq: skewness_h_0}) with (\ref{eq: skewness_FG1993}),  
we readily see the role of stochasticity to the skewness parameter 
as follows: while the width of scatter, $\epsilon_{\rm scatt}$, decreases 
the skewness amplitude, non-Gaussianity characterized 
by the third order moments, $\langle \epsilonh^2 \deltam \rangle$ and 
$\langle \epsilonh^3 \rangle$ conversely increases the overall amplitude. 
Indeed, this fact illustrates the general trend of the skewness parameters. 
That is, the skewness parameters can be schematically divided into the 
contributions from nonlinearity and stochasticity of biasing as follows:  
\begin{equation}
\label{eq:S_pt}
S_{\rm h}^{\rm (n)} =
\frac{1}{b_1}
\frac{\displaystyle \left( S_{\rm m}^{\rm (n)} + 3 \frac{b_2}{b_1} + 
[\mbox{3rd order stochasticity}] \right)}{\displaystyle 
\left(1 + \left[\mbox{2nd order stochasticity}\right] \right)}, 
\end{equation}
where the third order stochasticity characterizes the non-Gaussianity 
induced by the stochasticity of biasing and the second order 
stochasticity represents the degree of stochasticity.  
 
Equation (\ref{eq:S_pt}) illustrates the qualitative behavior
of the normalized skewness for the halo field: (i) it is inversely
proportional to the linear bias coefficient $b_1$, (ii) the nonlinearity
(with $b_2>0$) {\it increases} the skewness, and (iii) the stochasticity {\it
decreases} the overall amplitude of the skewness, but adds a positive
term to the purely gravitational term $S_{\rm m}$. 
Although it is not clear
which of the two opposite effects due to the stochasticity is more
important, the results in figures \ref{fig:genus_perturb_GA} 
and \ref{fig:skew_fit_G} show that 
total stochastic effect is not very strong
as far as those skewness parameters are concerned.
This indicates that two opposite effects due to the biasing
stochasticity on the skewness parameters [equation (\ref{eq:S_pt})]
are almost cancelled out by each other.

\end{document}